\documentclass[journal,onecolumn]{IEEEtran}

\usepackage{caption}

\usepackage{url}

\usepackage{amsmath,amsfonts,amsthm,mathtools}
\usepackage{graphicx}
\usepackage{bbm}
\usepackage{color}
\usepackage{blkarray}
\newtheorem{theorem}{Theorem}
\newtheorem{proposition}{Proposition}
\newtheorem{lemma}{Lemma}
\newtheorem{condition}{Condition}
\newtheorem{assumption}{Assumption}
\newtheorem{example}{Example}

\ifodd 0
\newcommand{\rev}[1]{{\color{blue}#1}} 
\newcommand{\com}[1]{\textbf{\color{red}(COMMENT: #1)}} 
\newcommand{\edt}[1]{\textbf{\color{magenta}#1}} 
\newcommand{\clar}[1]{\textbf{\color{green}(NEED CLARIFICATION: #1)}}

\else
\newcommand{\rev}[1]{#1}
\newcommand{\com}[1]{}
\newcommand{\edt}[1]{}
\newcommand{\clar}[1]{}
\fi

\begin{document}

\title{Using Private and Public Assessments in\\ Security Information Sharing Agreements}

\author{Parinaz~Naghizadeh,~\IEEEmembership{Member,~IEEE,}
            and Mingyan~Liu,~\IEEEmembership{Fellow,~IEEE}
\thanks{P. Naghizadeh is with the Department of Integrated Systems Engineering and Electrical and Computer Engineering at Ohio State University, email: \{naghizadeh.1@osu.edu\}. M. Liu is with the Department of Electrical Engineering and Computer Science at the University of Michigan, email: \{mingyan@umich.edu\}.}%
\thanks{To appear in the IEEE Transactions of Information Forensics and Security, 2020.}
}

\maketitle

\begin{abstract}
Information sharing among organizations has been gaining attention as a method for improving cybersecurity. However, the associated disclosure costs act as deterrents for firms' voluntary cooperation. In this work, we take a game-theoretic approach to understanding firms' incentives in these agreements. We propose the design of inter-temporal incentives (i.e. conditioning future cooperation on past interactions). 
Specifically, we show that incentives for full cooperation can be designed if firms share their private assessments of other firms' disclosure decisions through a common communication platform. We further show that similar incentives can be designed based on outcomes of a public rating/assessment system. 
\end{abstract}

\section{Introduction} \label{sec:intro}

Improving the ability of analyzing cyber-incidents, and ensuring that the results are shared among organizations and authorities in a timely manner, has received increased attention in recent years by governments and policy makers, as it can lead to better protection of the national infrastructure against potential cyber-attacks, allow organizations to invest in the most effective preventive and protective measures, and protect consumer rights.

In the US, improving information sharing was listed as one of President Obama's  administration's priorities on cybersecurity, 
as evidenced by its inclusion as a key focus area in Executive Order 13636 \cite{exec13636} on ``Improving Critical Infrastructure Cybersecurity''. 
During the first White House Summit on cybersecurity and consumer protection, Executive Order 13691: ``Promoting Private Sector Cybersecurity Information Sharing'' \cite{exec13691}, was signed to encourage companies to share cybersecurity information with one another and the federal government. Following the executive order, the Department of Homeland Security (DHS) started efforts to encourage the development of Information Sharing and Analysis Organizations  \cite{dhs-isao}, as well as the Cyber Information Sharing and Collaboration Program (CISCP) in order to encourage Cooperative Research and Development Agreements. 

In general, depending on the breach notification law or the information sharing agreement, a firm may be required to either publicly announce an incident, report it to other firms participating in the agreement, and/or notify affected individuals and appropriate authorities. Currenlty, existing laws in the US and the EU require organizations to report to an authority, with many also mandating notification of the affected individuals, e.g., HIPAA for the health sector in the US (see \cite{laube16} for a summary of prominent US and EU laws). However, motivated by the aforementioned trend in the newest initiatives in the US (in particular, E.O. 13691), in this paper, we are primarily interested in information sharing agreements \emph{among firms}, both with and without facilitation by an authority. Examples of existing agreements/organizations of this type include Information Sharing and Analysis Centers (ISACs), Information Sharing and Analysis Organizations (ISAOs), and the United States Computer Emergency Readiness Team (US-CERT). In particular, existing ISACs include the MS-ISAC  (for state, local, tribal, and territorial governments),  IT-ISAC (information technology sector), and FS-ISAC (financial services sector). %
Currently, joining and reporting in all such information sharing organizations is voluntary. 

\subsection{Problem Motivation}
Several studies have analyzed the positive effects of information sharing laws. Romanosky et al. \cite{romanosky11} show that the introduction of breach disclosure laws has resulted in a reduction in identity theft incidents. Gordon et al. \cite{Gordon15} argue that shared information can reduce the uncertainty in adopting a cybersecurity investment, thus leading firms to take a proactive rather than reactive approach to security, and consequently increasing the expected amount of investments in cybersecurity. %
Nevertheless, there exist anecdotal and empirical evidence that security breaches remain under-reported, see e.g., \cite{verizon08,threattrack}. %

These observed disincentives by companies for sharing security information, {both to an authority, as well as to other firms}, can be primarily explained by analyzing the associated economic impacts. Campbell et al. \cite{campbell03} and Cavusoglu et al. \cite{cavusoglu04} conduct event-study analyses of market reaction to breach disclosures, both demonstrating a drop in market values following the announcement of a security breach. In addition to an initial drop in stock prices, an exposed breach or security flaw can result in loss of consumer/partner confidence in a company, leading to a further decrease of revenues in the future.  
 Finally, documenting and announcing security breaches impose a bureaucratic burden on the company, e.g, when an agreement requires the reports to comply with a certain incident reporting terminology; examples include frameworks proposed by the DHS \cite{dhs-report} and Verizon's VERIS \cite{veris}. 

Given these potential disclosure costs, and the evidence of under-reporting of security information, it is clear that we need a better understanding of firms' incentives for sharing their security information. {Our focus on this paper is on firms' incentives for disclosure in information sharing agreements with other firms, as well as the economic incentives that could lead to voluntary cooperation by firms in these agreements.}

\subsection{Inter-temporal Incentives in Information Sharing}
In this paper, we take a game-theoretic approach to understand firms' behavior and (dis)incentives in security information sharing agreements {among firms}. This approach is motivated by the fact that despite the aforementioned disclosure costs (which deter firms from joining such agreements and sharing information {with one another}), disclosure has benefits for participating firms, as each firm can prevent similar attacks and invest in the best security measures by leveraging other firms' experience. Consequently, {when disclosure costs are not prohibitively high}, an outcome in which firms disclose their information would be both welfare maximizing and preferred by all participants. In other words, there is a conflict between individual interest and societal goals. To capture this conflict, we model security information sharing agreements as an N-person prisoner's dilemma (NPD) game. In an NPD, there is no information sharing at the equilibrium, as also predicted by similar game-theoretic models which consider one-shot information sharing games (see Section \ref{sec:related}). Existing research has further proposed audits and sanctions (e.g. by an authority or the government), or introducing additional economic incentives (e.g. taxes and rewards for members of ISACs) as remedies for encouraging information disclosure.

In this paper, we take a different approach and account for the ongoing, repeated nature of these agreements to propose the design of inter-temporal incentives that lead sufficiently patient firms to cooperate on information sharing. It is well known in the economic literature that repetitions of an otherwise non-cooperative and inefficient game can lead economically rational agents to coordinate on efficient equilibria; see \cite{mailath06}. The possibility of achieving efficient outcomes however depends on whether the monitoring of other participants' actions is perfect or imperfect, and private or public. In particular, for information sharing games, each firm or an outside monitor can (at best) only imperfectly assess the  comprehensiveness of the shared information. Accordingly, we consider two possible monitoring structures. 

First, we consider the design of cooperation incentives when firms have access to a communication platform, through which they can report their private beliefs on whether other firms are adhering to the agreement. We show that given a simple imperfect private monitoring structure by each firm, the folk theorem of \cite{kandori98} is applicable to our proposed NPDs, thereby incentivizing information sharing.  
We then analyze the role of a rating/assessment entity in providing an imperfect public signal about the quality of firms' reports in the agreement. We show that the folk theorem of \cite{fudenberg94} is applicable in this scenario, therefore again making it possible to design appropriate cooperation incentives. We illustrate the construction of such incentives through an example, and discuss the effect of the monitoring accuracy on the construction.
We also discuss some practical implications of these findings. 

\subsection{Contributions and Paper Organization}
The contributions of this paper are as follows:

$\bullet$ We propose the design of inter-temporal incentives for supporting cooperative behavior in security information sharing agreements. To this end, we model firms' interactions as an N-person prisoner's dilemma game equipped with a simple monitoring structure. 

$\bullet$ We analyze the effectiveness of sustaining cooperation through inter-temporal incentives by proposing a platform for communication through which firms share their private assessments of one another. 

$\bullet$ We analyze the effectiveness of using a public rating/assessment system in providing imperfect public monitoring to sustain similar incentives, and illustrate the construction of such incentives through a numerical example.  

\rev{$\bullet$ We discuss the potential practical and policy implications of these findings in the design and operation of security information sharing agreements.}

Preliminary versions of this work appeared in \cite{naghizadeh16a,naghizadeh16b}. We first proposed the idea of using inter-temporal incentives in information sharing agreements in \cite{naghizadeh16a}, and analyzed the possibility of using \rev{\emph{public}} monitoring in a two-person prisoner's dilemma game in \cite{naghizadeh16b}. \rev{In this paper, we establish the possibility of using \emph{private} monitoring along with a communication platform to design inter-temporal incentives. We further generalize the model to N-person prisoner's dilemma games, show that our analysis of public monitoring in \cite{naghizadeh16b} extends to this new model, illustrate the construction of incentives using a numerical example, and discuss practical implications.}

\emph{Paper Organization:}
We present the model and proposed monitoring scheme for information sharing games in Section \ref{sec:model}. We discuss the feasibility of providing inter-temporal incentives for cooperation using private observations and communication among firms in Section \ref{sec:private}, followed by the role of a public monitoring system and a numerical example in Section \ref{sec:public}. {We discuss practical aspects in Section \ref{sec:disc},} followed by related work in Section \ref{sec:related}. Section \ref{sec:conclusion} concludes the paper.

\section{Information Sharing Game Model} \label{sec:model}

{We first formally introduce the information sharing game model among firms. We begin by presenting the one-shot, single stage information sharing game in Section \ref{sec:stagegame}, followed by the extensions required to study the repeated game, namely, the monitoring structure, in Section \ref{sec:repeatedgame}. A list of notation used in the model is given in Table \ref{t:notations}.} 

\subsection{The Stage Game}\label{sec:stagegame}

Consider $N$ (symmetric) firms participating in an information sharing agreement (e.g. firms within an ISAC). Each firm can choose a level of expenditure in security measures to protect her infrastructure against cyber incidents. Examples include implementing an intrusion detection system, introducing employee education initiatives, and installing and maintaining up-to-date security software. {We assume these measures are implemented independently of other firms' security expenditure decisions, i.e., we do not model positive/negative externalities of firms' security investments on one another. This allows us to focus solely on firms' information sharing decisions.}\footnote{{Specifically, the information shared by firm $i$ may be a \emph{substitute} to firm $j$'s investment, i.e., firm $j$ may decrease her security expenditure when she receives information from firm $i$. When firms' security decisions are interdependent due to risk spillovers, this potential reduction in positive externalities from $j$'s investments would result in firm $i$ revising both her security expenditure level as well as her disclosure decisions, obscuring the incentives for information sharing. We therefore remove these effects by decoupling the security expenditure decisions. Note that while we assume independent security expenditures among firms, the security decision of each firm can still be affected by others' information disclosure decisions. Analyzing the simultaneous interplay of investment and sharing decisions remains a direction of future work.}}  

\subsubsection{Firms' Actions} The information sharing agreement requires each firm $i$ to share her security information with other participating firms. This can include information on both successful and failed attacks, as well as effective breach prevention methods and the firm's adopted security practices. A firm $i$ should decide whether to fully disclose such information. We denote the decision of firm $i$ by $r_i\in \{0,1\}$, with $r_i=0$ denoting (partially) concealing and $r_i=1$ denoting (fully) disclosing.\footnote{The results and intuition obtained in the following sections continue to hold when firms can choose one of finitely many disclosure levels, given an appropriate extension of utilities and the monitoring structure.} Denote the number of firms adopting a full disclosure decision by $x$; i.e., $x:=|\{i  | \ r_i=1\}|$. 

A decision of $r_i=1$ is beneficial for the following reasons. First, other firms $j\neq i$ can leverage the disclosed information to protect themselves against ongoing attacks and to adopt better security practices. In addition, the disclosed information may provide a competitive advantage to firms $j\neq i$, allowing a firm $j$ to increase her share of the market by strategically leveraging the attained information to attract a competitor $i$'s customers. Further, sharing of security information may be beneficial to firm $i$ herself as well (especially when several other firms are disclosing), as it may garner trust from potential partners and customers. 
We denote all such applicable \emph{information gains} to a firm, as a function of firm $i$'s decision and the number of \emph{other} firms making a full disclosure decision, by $G(r, z): \{0,1\}\times\{0,1, \ldots, N-1\}\rightarrow \mathbb{R}_{\geq 0}$, with $G(0,0)=0$. We assume that given $r$, $G(r, \cdot)$ is increasing in $z$, the number of other firms disclosing.  

Despite the aforementioned benefits of adopting $r_i=1$, firm $i$ has a disincentive for full disclosure due to the associated costs. These costs includes the man-hours spent in documenting and reporting security information, as well as potential losses in reputation, business opportunities with potential collaborators, stock market prices, and the like, following the disclosure of a breach or existing security flaws. 
In addition, it may be in $i$'s interest to conceal methods for preventing ongoing threats, predicting that an attack on the competitor $j$ will result in $j$'s customers switching to $i$'s products/services, increasing firm $i$'s profits. Consequently, such potential market loss or competitor's gain in sales can further deter firms from adhering to information sharing agreements. We denote all these associated \emph{disclosure costs} by $L(r, z): \{0,1\}\times\{0, \ldots, N-1\}\rightarrow \mathbb{R}^{+}$, with $L(0,0)=0$, where the cost can potentially depend on how many other firms, $z$, are disclosing their security information. 

\subsubsection{Firms' Utilities} 
{Let $u_i(\mathbf{r})$ denote the utility of firm $i$ at the profile of disclosure decisions $\mathbf{r}=\{r_1, \ldots, r_N\}$. We assume that, given the disclosure gains and costs defined above, firms' utilities are given by 
\begin{align}
u_i(\mathbf{r}):=G(r_i, x- \mathbbm{1} \{r_i=1\}) - L(r_i, x- \mathbbm{1} \{r_i=1\}),
\label{eq:utilities}
\end{align}
where $x$ is the number of firms adopting full disclosure decisions, and $\mathbbm{1}(\cdot)$ denotes the indicator function. 
In particular, substituting for each firm's disclosure decision, we get the following utilities for the cooperators ($r_i=1$) and deviators ($r_i=0$):} 
\begin{align*}
\text{Cooperator:} & \qquad C(x) := G(1,x-1) - L(1,x-1)~,\\
\text{Deviator:} & \qquad D(x) := G(0,x) - L(0,x)~.
\end{align*}

We impose the following two assumptions on these functions: 
\begin{assumption}
Non-cooperation dominates cooperation: 
\begin{align*}
\text{(A1)} &\qquad D(x-1) > C(x),~ \forall 1\leq x\leq N~.
\end{align*} 
\end{assumption}
Assumption (A1) entails that the disclosure costs outweigh the gain from sharing for the firm, making $r_i=0$ a dominant strategy. In other words, the marginal benefit from increased trust or approval due to disclosure (if any) is limited compared to the potential market and reputation loss due to disclosed security weaknesses. Therefore, the only Nash equilibrium of a one-shot information sharing game is for no firm to disclose her information. This observation is consistent with similar studies of one-shot information sharing games in \cite{laube16,gordon03}. 

\begin{assumption}
Non-cooperation is inefficient: 
\begin{align*}
\text{(A2)} &\qquad C(N) > D(0) = 0~.
\end{align*} 
\end{assumption}
Assumption (A2) entails that the resulting non-disclosure equilibrium is suboptimal, particularly compared to the outcome in which all firms disclose. That is, full disclosure dominates the unique Nash equilibrium of the one-shot game. We may further be interested in imposing a more restrictive condition (although this is not necessary for our technical discussion).  
\begin{align*}
\text{(A2')} &\quad xC(x) + (N-x)D(x) > \notag\\
& \qquad(x-1)C(x-1) + (N-x+1)D(x-1),~ \forall x.
\end{align*} 
Under (A2') (which indeed implies (A2)), non-disclosure by any firm decreases social welfare, making the full disclosure equilibrium $x=N$ the socially desired outcome.  

\subsubsection{The Security Information Sharing Game} {We refer to the game  $<\{1,\ldots,N\}, \{r_i\}, \{u_i\}>$ as the security information sharing game;  it consists of the $N$ firms as players, taking actions $r_i\in\{0,1\}$, and receiving utilities $u_i$ given in \eqref{eq:utilities}. 
In particular, when the firms' utilities satisfy assumptions (A1) and (A2) (or (A2')), the security information sharing game will be an instance of the $N$-person Prisoner's Dilemma (NPD) game;} see e.g., \cite{bonacich76,goehring76}. 
These games model situations in which there is a conflict between individual and societal goals, e.g., individual decisions whether to belong to unions, political parties, or lobbies, and problems of pollution or overpopulation \cite{bonacich76}. The imposed assumptions then model the intuition that in such situations, any individual has a disincentive for cooperation, (A1), despite the fact that an outcome in which all  cooperate would have been preferred by each participant, (A2). {Below, we provide two examples of firms' utility functions that satisfy these assumptions.} 

\begin{example}
Consider the gain functions $G(1,z)=G(0,z)=zG$ and loss functions $L(0,z)=0$ and $L(1,z)=L$. Here, each firm obtains a constant gain $G$ from any other firm who is disclosing information, and incurs a constant loss $L$ if she discloses herself, both regardless of the number of other firms making a disclosure decision. It is easy to verify that these functions satisfy (A1). Furthermore, if $G>\frac{L}{N-1}$, assumptions (A2) and (A2') hold as well. Note also that the 2-player prisoner's dilemma can be recovered as a special case when $N=2$. \hfill $\qed$
\end{example}

\begin{example}
Alternatively, consider the gain functions $G(1,z)=G(0,z)=f(z)G$, where $f(\cdot):\{0,\ldots, N-1\}\rightarrow \mathbb{R}^{+}$ is an increasing and concave function, and loss functions $L(0,z)=0$ and $L(1,z)=L$. The concavity of $f(\cdot)$ implies that as the number of cooperators increases, the marginal increase in information gain is decreasing due to potential overlap in the disclosed information. The utilities of cooperators and deviators will be given by: 
\begin{align*}
C(x) = f(x-1)G - L, \text{ and, } D(x) = f(x)G~.
\end{align*}
Assumption (A1) follows. Assumptions (A2) will hold if and only if $G>\frac{L}{f(N-1)-f(0)}$. However, unlike the previous example, for (A2') to hold we need additional restrictions beyond that required for (A2). Specifically, the full disclosure equilibrium will be the optimal solution only if the constants $G$ and $L$ are such that: 
\begin{align*}
&(N-x)\left(f(x)-f(x-1)\right)\notag\\
&~~ +(x-1)\left(f(x-1)-f(x-2)\right) >\frac{L}{G},~ \forall x. \qquad \hfill \qed
\end{align*}
\end{example}

{As evidenced by these two examples, the general utility models of NPDs can capture \rev{the classic 2-player prisoner's dilemma games as a special case (Example 1), while also capturing conflicts beyond those modeled by the 2-player prisoner's dilemma game  (Example 2).}} 

\begin{table}
\caption{Summary of Notation}
\label{t:notations}
\centering
{
\begin{tabular}{|c|p{0.75\columnwidth}|} 
 \hline
 Symbol & Description \\
 \hline
$N$ & Number of firms\\ 
\hline 
$r_i$ & Disclosure actions selected from $\{0,1\}$\\
\hline
$\mathbf{r}$ & Profile of all firms' disclosure actions\\
\hline
$x$ & Number of firms adopting a full disclosure decision\\
\hline
$z$ & Number of firms other than a given firm who disclose information\\
\hline
$G(r,z)$ & Benefit from information sharing to a firm\\
\hline
$L(r,z)$ & Loss from information sharing to a firm\\
\hline
$u_i(\mathbf{r})$ & Utility functions given in \eqref{eq:utilities}\\
\hline
$C(x)$ & Utility attained by the cooperators (i.e., $r_i=1$) as a function of the number of  reporting firms\\
\hline
$D(x)$ & Utility attained by the deviators (i.e., $r_i=0$) as a function of the number of  reporting firms\\
\hline
$b_{ij}$ & Belief of firm $i$ about the action $r_j$ of firm $j$\\
\hline
$\mathbf{b}$ & Profile of beliefs observed by all firms/the monitor\\
\hline
$\pi(b_{ij}|r_{j})$ & Distribution of firm $i$'s belief about firm $j$'s action\\
\hline
$\epsilon$ & Probability that a firm $i$/the monitor can detect security information not known by firm $j$\\
\hline
$\alpha$ & Probability that a firm $i$/the monitor can detect security information not reported by firm $j$\\
\hline
\end{tabular}
}
\end{table}

\subsection{Repeated Interactions and the Monitoring Structure}\label{sec:repeatedgame}
{When the information sharing game described above is played only once, the unique Nash equilibrium will be to share no information (i.e., $r_i=0, \forall i$). We are alternatively interested in considering \emph{repeated} security information sharing games, and in leveraging the repeated nature of these agreements in the design of inter-temporal incentives that can lead firms to adopt full disclosure decisions at each stage}. Such inter-temporal incentives should be based on the history of firms' past interactions. We therefore formalize firms'  {capabilities in monitoring others' actions, which can be used to construct the histories of past interactions.}

First, note that such monitoring is inevitably \emph{imperfect}; after all, the goal of an information sharing agreements is to encourage firms to reveal their unverifiable and private breach and security information. Furthermore, the monitoring can be either carried out independently by the firms, or be based on the reports of a central monitoring system. We consider both possibilities.  

\subsubsection{Imperfect Private Monitoring}
First, assume each firm conducts her own monitoring and forms a belief on other firms' disclosure decisions. Specifically, by monitoring firm $j$'s externally observed security posture, firm $i$ forms a \emph{belief} $b_{ij}$ about $j$'s report. We let $b_{ij}=1$  indicate a belief by firm $i$ that firm $j$ has fully disclosed all information, and $b_{ij}=0$ otherwise. 
In other words, $b_{ij}=0$ indicates that firm $i$'s monitoring provides her with evidence that firm $j$ has experienced an undisclosed breach or has an unreported security flaw. 
Formally, we assume the following distribution on firm $i$'s belief given firm $j$'s report: 
\begin{align}
\pi(b_{ij} | r_j) = \left\{\begin{array}{lr}
        \epsilon, & \text{for } b_{ij}=0, r_j=1\\    
        1 - \epsilon, & \text{for } b_{ij}=1, r_j=1\\    
        \alpha, & \text{for } b_{ij} = 0, r_j=0\\
        1 - \alpha, & \text{for } b_{ij}=1, r_j=0\\
        \end{array}\right.
        \label{eq:private-mon}
\end{align}
with $\epsilon \in (0,1/2)$ and $\alpha \in (1/2,1)$. First, note that $\epsilon$ is in general assumed to be small; that is, if firm $j$ fully discloses all information ($r_j=1$), firm $i$'s belief will be almost consistent with the received information. Intuitively, this entails the assumption that with only a small probability $\epsilon$, firm $i$ will be observing flaws or breaches that have gone undetected by firm $j$ herself, as internal monitoring is more accurate than externally available information. On the other hand, firm $i$ has an accuracy $\alpha$ in detecting when firm $j$ conceals security information ($r_j=0$). Note that ($\epsilon=0, \alpha=1$) is equivalent to the special case of perfect monitoring.  

We assume the evidence available to firm $i$, and hence the resulting belief $b_{ij}$, is private to firm $i$, and independent of all other beliefs. Specifically, $b_{ij}, \forall i\neq j$ are i.i.d. samples of a Bernoulli random variable (with parameter $\alpha$ or $\epsilon$ depending on $r_j$).

\subsubsection{Imperfect Public Monitoring} Alternatively, consider an independent entity (the government, a white hat, or a research group), {commonly agreed on by the members of the information sharing agreement}, and 
referred to as \emph{the monitor}, who assesses the comprehensiveness of firms' disclosure decisions, and publicly announces the results. We assume the distribution of the beliefs $\{b_{01}, \ldots, b_{0N}\}$ formed by the monitor is:
\begin{align}
\hat{\pi}(\{b_{01}, \ldots, b_{0N}\} | \{r_{1}, \ldots, r_{N}\}) := \Pi_{j=1}^N \pi(b_{0j} | r_j)~,
        \label{eq:public-mon}
\end{align}
where the distributions $\pi(b_{0j} | r_j)$ follow \eqref{eq:private-mon}, with $\epsilon$ and $\alpha$ interpreted similarly. Note that the monitoring technology of the monitor, i.e. $(\alpha, \epsilon)$, may in general be more accurate than that available to the firms. 
{It is worth mentioning that the binary beliefs are assumed for ease of exposition; the results of the subsequent sections continue to hold if the monitoring technology has finitely many outputs.}

\section{Imperfect Private Monitoring: The Role of Communication} \label{sec:private}

In this section, we consider the use of private monitoring in providing inter-temporal incentives for information sharing. \rev{We ask whether it is possible for firms' to establish collaborative information sharing agreements in the long-run when each firm can only privately monitor and form beliefs on others' disclosure decisions.} In particular, we are interested in a folk theorem for the information sharing game of Section \ref{sec:model}; a folk theorem is a full characterization of payoffs that can be achieved as average payoffs of the infinitely repeated game when firms are sufficiently patient. \rev{If a folk theorem exists and is applicable to our model, it will establish the possibility that the simple inter-firm monitoring structure \eqref{eq:private-mon} can be leveraged to incentivize long-run collaboration among firms.}

However, unlike repeated games with public monitoring, relatively little is known about games with private monitoring. In particular, \cite{fudenberg86} and \cite{fudenberg94} present folk theorems under perfect and imperfect \emph{public} monitoring, respectively, by requiring relatively general conditions on the underlying game and monitoring technology. The possibility of these results hinges heavily on that firms share common information on each others' actions (i.e., the public monitoring outcome), as a result of which it is possible to recover a recursive structure for the game; this gives rise to the folk theorem. However, a similar folk theorem with private monitoring remained an open problem until recently,\footnote{A recent advance in the field is by Sugaya \cite{sugaya13}, who presents a folk theorem for repeated games with imperfect private monitoring, without requiring cheap talk communication or public randomization. The conditions on the private monitoring structure required by Sugaya's folk theorem are however more restrictive than those of Kandori and Matsushima's \cite{kandori98} folk theorem with (cheap talk) communication used in this section. Therefore, we analyze the application of the folk theorem with communication of \cite{kandori98}; this will further allow us to draw a closer parallel with the public monitoring structure discussed later in Section \ref{sec:public}.} mainly due to the lack of a common public signal. Nevertheless, the possibility of cooperation, and in particular folk theorems, have been shown to exist for some special classes of such games. Examples include games in which firms are allowed to communicate (cheap talk) \cite{compte98,kandori98}, those in which firms have public actions (e.g., announcement of sanctions) \cite{park11}, and games with almost public monitoring, i.e., private monitoring with signals that are sufficiently correlated \cite{mailath02}.

\rev{Below, in Section \ref{sec:folk-private}, we present one such folk theorem, with private monitoring and communication, due to \cite{kandori98}, and  in Section \ref{sec:verify-private}, verify that it applies to NPD information sharing games with monitoring given by \eqref{eq:private-mon}.} We will elaborate on the practical and policy implications of this finding in Section \ref{sec:disc}. 

\subsection{The Folk Theorem with Imperfect Private Monitoring and Communication} \label{sec:folk-private}
{
 At the stage game, each firm $i$ chooses a disclosure action $r_i\in R_i$, leading to the profile of actions $\mathbf{r} \in R:=\prod_{i=1}^N R_i$. Following the choice of actions, each firm privately observes an outcome $b_i \in B_i$ through her monitoring of other firms, where $B_i$ is a finite set of possible signals. {For the information sharing game of Section  \ref{sec:model},   $R_i=\{0,1\}$ and $B_i=\{0,1\}^{N-1}$.} 
 
 The probability of observing the profile of private signals $\mathbf{b}\in B:=\prod_{i=1}^N B_i$ following $\mathbf{r}$ is given by the joint distribution $\pi(\mathbf{b}|\mathbf{r})$ (e.g., the joint distribution of the private monitoring technologies in \eqref{eq:private-mon}). Assume $\pi$ has full support, i.e., $\pi(\mathbf{b}|\mathbf{r})>0, ~\forall \mathbf{b}, \forall\mathbf{r}$. Let $u_i^*(r_i, b_i)$ be the utility of firm $i$ when she plays $r_i$ and observes the signal $b_i$. Note that $i$'s utility depends on others' actions only through $b_i$, and thus the stage payoffs are not informative about others' actions. The ex-ante stage game payoff for firm $i$ when $\mathbf{r}$ is played is therefore given by   
$u_i(\mathbf{r}) = \sum_{\mathbf{b}\in B} u_i^*(r_i, b_i)\pi(\mathbf{b}|\mathbf{r})$.\footnote{Alternatively, we can fix the ex-ante payoffs of \eqref{eq:utilities} as the model primitives, and consider various monitoring technologies according to \eqref{eq:private-mon}; payoffs $u_i^*$ can be adjusted accordingly.}
%

In order to construct an equilibrium with communication, we allow firms to make announcements at each stage of this game. Formally, after choosing the action $r_i$ and observing the signal $b_i$, each firm $i$ will publicly announce a message $m_i\in M_i$, selected from the finite set of possible messages $M_i$. Let $M=\Pi_{i=1}^N M_i$ denote the space of all possible messages, {which can in general include firms' actions and/or observations.} 
The strategy $s_i=(r_i, m_i)$ of a firm at each stage game will consist of both an action $r_i$ and a message $m_i$. \rev{We will later choose each firm's private belief $b_i$ as her message.}

Given the above stage game, we now discuss firms' strategies in the repeated game. In the infinitely repeated game, the strategy $s_i$ specifies firm $i$'s actions and messages for each time step $t$, i.e, $s_i:=(s_i(t))_{t=0}^{\infty} = (r_i(t), m_i(t))_{t=0}^{\infty}$, where:
\begin{align*}
& r_i(t): R_i^{t-1}\times B_i^{t-1} \times M^{t-1} \rightarrow \Delta(R_i)~,\\
& m_i(t): R_i^{t}\times B_i^{t} \times M^{t-1} \rightarrow \Delta(M_i)~.
\end{align*}
Let $r_i^t=(r_i(0), \ldots, r_i(t))$ be the profile of firm $i$'s actions up to some finite time $t$. Define $b_i^t$ and $m^t$ similarly. {Note that the domain of the firm's strategies consists of two types of history: a \emph{private history} $h_i^t:=(r_i^{t}, b_i^{t})$ containing her own past actions and beliefs about others' actions, as well as a \emph{public history} $h^t:=m^t$ of the messages communicated so far.} The strategy $s^t_i=(r_i^t, m_i^t)$ of firm $i$ at time $t$ is in general based on both the public and private histories, i.e., it is a mapping from $(h_i^{t-1}, h^{t-1})$ to (a probability distribution over) $R_i$ determining her next play. 

Given the strategy profiles $\mathbf{s}=(s_1, \ldots, s_N)$, and assuming that firms discount future payoffs by a discount factor $\delta$, a firm's average payoff throughout the repeated game is given by $v_i(\mathbf{s}, \delta):=(1-\delta)\sum_{t=0}^\infty \delta^t \mathbb{E} [u_i(r_i(t), b_i(t))|~\mathbf{s}]$. Each firm $i$ is choosing her strategy $s_i$ to maximize the expected value of this expression. 

\paragraph*{Equilibrium concept} We are interested in characterizing the payoffs attainable by the strategy profiles $\mathbf{s}$ that are a \emph{sequential equilibrium} of the game. Formally, $\mathbf{s}$ is a sequential equilibrium of the game if for every firm and every history of the firm $(h_i^t, h^t)$, the strategy selected by the firm, $s_i|_{(h_i^t, h^t)}$, is a best reply to $E[s_{-i}|_{h^t_{-i}, h^t}|h_i^t]$, which is the belief of firm $i$ over other firms' strategies, given her private history. That is, a firm is best-responding according to her belief over private histories of other firms, in particular those which are consistent with her own private history (see also, \cite[Definition 12.2.3]{mailath06}).

\paragraph*{The folk theorem} 
{Assume firms are interested in achieving a payoff profile $\{v_i(\mathbf{s}, \delta)\}_{i=1,\ldots, N}$ as the sequential equilibrium of the repeated game. In order to reach these expected payoffs, at each stage $t$, firms should select some profile of actions $\mathbf{r}^t$; \rev{for instance, this could be the profile of actions $r^t_i=1, \forall i$, to achieve full cooperation}. To ensure that firms do indeed follow this payoff profile, they need to be able to detect and appropriately punish deviations. \rev{That is, firms should be able to (collectively) determine if any firm is not disclosing her information, and reduce  their own information sharing or terminate the ongoing agreement in response.} 
The main goal of the folk theorem is to identify conditions under which this is possible; that is, conditions on the firms' private monitoring accuracy and their actions in the communication stage, through which a given payoff profile can emerge as a sequential equilibrium of the repeated game.
%
%
More generally, folk theorems identify conditions under which it is possible to construct sequential equilibria which can achieve \emph{any} payoff in (the interior of) the set $\mathcal{F}^*$ of feasible and strictly individually rational payoffs of the game}\footnote{Formally, $\mathcal{F}^*$ is defined as follows. Let $\mathcal{F}^\dagger$ denote the set of convex combinations of firms' payoffs for outcomes in $R$, i.e., the convex hull of $\{(u_1(\mathbf{r}), \ldots, u_n(\mathbf{r}))| \mathbf{r}\in R\}$. We refer to $\mathcal{F}^\dagger$ as the set of \emph{feasible} payoffs. Of this set of payoffs, we are particularly interested in those that are \emph{individually rational}: an individually rational payoff profile $\mathbf{v}$ is one that gives each firm $i$ at least her minmax payoff $\underline{v}_i:=\min_{\boldsymbol{\rho}_{-i}}\max_{r_i} u_i(r_i, \boldsymbol{\rho}_{-i})$, where $\boldsymbol{\rho}_{-i}$ denotes a mixed strategy profile by firms other than $i$. Formally, let $\boldsymbol{\rho}^i$, with
$\boldsymbol{\rho}_{-i}^i:= ~\arg\min_{\boldsymbol{\rho}_{-i}}\left(\max_{r_i} ~ u_i(r_i, \boldsymbol{\rho}_{-i})\right)~,~~
{\rho}_i^i:= ~\max_{r_i} ~~ u_i(r_i, \boldsymbol{\rho}_{-i}^i)~,$
denote the minmax profile of firm $i$. Then, $\mathcal{F}^*:=\{\mathbf{v}\in \mathcal{F}^\dagger| v_i > \underline{v}_i, \forall i\}$ will be the set of feasible and strictly individually rational payoffs.} for $\delta$ sufficiently close to 1. It is worth mentioning that the set $\mathcal{F}^*$ contains the set of Pareto efficient payoffs of the game. Therefore, a folk theorem states that one can construct a sequential equilibrium of the game through which firms can achieve arbitrarily efficient payoffs.

Specifically, assume that at the end of each stage $t$, each firm $i$ is asked to report her privately observed signal as her message, i.e., $m_i(t)=b_i(t)$. To make sure that firms truthfully report their signals, the equilibrium strategies use this private information solely to determine \emph{other} firms' deviations and future payoffs, and maintain $i$'s payoff independent of her report. As a result, truthful reporting of privately observed signals will be a (weak) best-response.\footnote{It is also possible to make truth reporting a \emph{strict} best-response if firms' privately observed signals are mutually correlated; see \cite[Section 4.2]{kandori98}.}$^,$\footnote{Note that unlike the public strategies played under public monitoring (Section \ref{sec:public}), each firm will be playing a private strategy at equilibrium, as she is using her private information in her message $m_i$. However, the choice of action $r_i$ will still be based only on the public information, i.e., the disclosed messages available to all firms.} It remains to ensure that the available signals are sufficiently informative: the signals should be distributed such that they allow firms to statistically distinguish between deviations by two different firms, as well as different deviations by the same firm. We now formally specify these conditions. 

First, define the following vectors: 
\begin{align*}
p_{-i}(\mathbf{r}) &:= (\pi_{-i}(\mathbf{b}_{-i}|\mathbf{r}))_{\mathbf{b}_{-i}\in B_{-i}}~, \\
\rev{p_{-\{i,j\}}}(\mathbf{r}) &:= (\rev{\pi_{-\{i,j\}}}(\rev{\mathbf{b}_{-\{i,j\}}}|\mathbf{r}))_{\mathbf{b}_{-ij}\in B_{-ij}}~\\
Q_{ij}(\mathbf{r}) &:= \{\rev{p_{-\{i,j\}}}(\mathbf{r}_{-i}, r_i')| r_i' \in R_i\backslash\{r_i\}\}~, 
\end{align*}
where $B_{-i}:=\Pi_{k\neq i}B_k$, $\rev{B_{-\{i,j\}}}:=\Pi_{k\neq i,j}B_k$, and $\pi_{-i}$ and \rev{$\pi_{-\{i,j\}}$} are marginal distributions of the joint distribution $\pi(\mathbf{b}|\mathbf{r})$ of privately observed signals. {In words, $p_{-i}(\mathbf{r})$ is the distribution of the private beliefs of firms other than $i$ under action profile $\mathbf{r}$. The interpretation for $\rev{p_{-\{i,j\}}}(\mathbf{r})$ is similar. Lastly, $Q_{ij}(\mathbf{r})$ is the distribution of the beliefs of firms other than $i$ and $j$, under deviations of firm $i$ from the profile $\mathbf{r}$.} \rev{Together, these belief vectors capture the collective information available to all firms about others' potential deviations, once all private beliefs are publicly announced through the agreement's communication platform. These beliefs need to be sufficiently informative, in a sense described below, so that the firms can maintain collaboration in the long-run.}
Specifically, the three sufficient conditions on the informativeness of signals required for the folk theorem to hold can be expressed using these three vectors, and are given below. 

\begin{condition}
At the minmax strategy profile of a firm $i$, ${\boldsymbol{\hat\rho}}^i$, for any firm $j\neq i$ and any mixed strategy $\rho'_j \in \Delta(R_j)$, either
\begin{align*}
(i) &\quad p_{-j}({\boldsymbol{\hat\rho}}^i) \neq p_{-j}({\boldsymbol{\hat\rho}}_{-j}^i, \rho'_j) ~~\text{or}, \\
(ii) & \quad p_{-j}({\boldsymbol{\hat\rho}}^i) =  p_{-j}({\boldsymbol{\hat\rho}}_{-j}^i, \rho'_j) \text{ and } u_j({\boldsymbol{\hat\rho}}^i)\geq u_j({\boldsymbol{\hat\rho}}_{-j}^i, \rho'_j). 
\end{align*}
\end{condition}

Condition (C1) states that at the minmax profile of any firm, a deviation by another firm is either statistically distinguishable (part (i)), and if not, it reduces the payoff of the deviator, and is hence not profitable (part (ii)). \rev{This assumption ensures that we can provide incentives to firms to punish (minmax) one another.} 

\begin{condition}
For each pair of firms $i\neq j$, and each pure action equilibrium $\mathbf{r}$ leading to an extreme point of the payoff set $\mathcal{F}^\dagger$, we have:
\[\rev{p_{-\{i,j\}}}(\mathbf{r}) \notin co(Q_{ij}(\mathbf{r})\cap Q_{ji}(\mathbf{r}))~,\]
where $co(X)$ denotes the convex hull of the set $X$. 
\end{condition}

Recall that $Q_{ij}(\mathbf{r})$ denotes the vector of distribution of beliefs of firms other than $i$ and $j$, when firm $i$ is deviating. \rev{(C2) therefore requires that a deviation by either $i$ or $j$ (but not both) is statistically detected by the remaining firms.}  

\begin{condition} 
For each pair of firms $i\neq j$, and each pure action equilibrium $\mathbf{r}$ leading to an extreme point of the payoff set $\mathcal{F}^\dagger$, we have:
\[co(Q_{ij}(\mathbf{r})\cup \rev{p_{-\{i,j\}}}(\mathbf{r})) \cap co(Q_{ji}(\mathbf{r})\cup  \rev{p_{-\{i,j\}}}(\mathbf{r})) = \{\rev{p_{-\{i,j\}}}(\mathbf{r})\}~.\]
\end{condition}

\rev{Finally, (C3) requires that firms other than $i,j$ can statistically distinguish deviations by $i$ from deviations by $j$, as the resulting distribution on $\rev{\mathbf{b}_{-\{i,j\}}}$ will be different under either firm's deviation. In other words, the only consistent distribution arises when neither firm is deviating.} 

Therefore, given adequate private monitoring signals and communication, we have the following folk theorem under imperfect private monitoring.

\begin{theorem} {\sc{(The Imperfect private monitoring with communication folk theorem \cite{kandori98}}).}\label{thm:private}
Assume that there are more than two firms ($N>2$), and the set of feasible and strictly individual rational payoffs $\mathcal{F}^*\subset \mathbb{R}^N$ has non-empty interior (and therefore dimension $N$). Then, if the monitoring of firms satisfy conditions (C1), (C2), and (C3), any interior payoff profile $\mathbf{v}\in \text{int}\mathcal{F}^*$ can be achieved as a sequential equilibrium average payoff profile of the repeated game with communication, when $\delta$ is close enough to 1. 
\end{theorem}

\subsection{Cooperation in Information Sharing Agreements}
\label{sec:verify-private}
We now show that the above folk theorem holds in the information sharing games with imperfect private monitoring structure given by \eqref{eq:private-mon}. That is, when the firms are sufficiently patient, they can sustain cooperation on any desired feasible and individually rational payoff, and in particular, the full security information sharing in a repeated setting, by truthfully revealing their private signals, and making their disclosure decisions based only on the imperfect, publicly announced collective observation about their past actions. 

To this end, we need to verify that the three conditions of the folk theorem on the informativeness of the monitoring signals hold for the joint distribution of the private signals in \eqref{eq:private-mon}. This is indeed true as shown in the following lemma; the proof is given in the appendix. 

\begin{lemma}\label{lemma:private}
The conditions (C1), (C2) and (C3) of the folk theorem in Theorem \ref{thm:private} hold with private monitoring distributions given in \eqref{eq:private-mon}. 
\end{lemma}

{The intuition behind the proof is as follows. Once firms' private beliefs are truthfully reported, it is as if we have access to $N-1$ independent realizations of the distribution in \eqref{eq:private-mon}. That is, as the signal distributions of the non-deviators are identical, to test the conditions of the folk theorem, it is sufficient to randomly choose one of the available cross-observations about a possible deviator (from firms other than the suspect $i$ for verifying (C1), or other than the two suspects $i,j$ for verifying (C2) and (C3)) and test the statistical distinguishability of that signal as required by the conditions (C1)-(C3).} %
Lemma \ref{lemma:private} together with Theorem \ref{thm:private} therefore establish the following.} 

\begin{proposition}
When firms are sufficiently patient (i.e., place high value on the future outcomes of their information sharing agreement), use private monitoring \eqref{eq:private-mon}, and are allowed to communicate their private signals, it is possible for them to nearly efficiently cooperate on full information disclosure through repeated interactions.
\end{proposition}

\section{Imperfect Public Monitoring: The Role of Centralized Monitoring} \label{sec:public}

{The possibility of public monitoring (either perfect or imperfect) can also enable the design of inter-temporal incentives for cooperation. In particular, instead of coordinating on their announced private observations (as in Section \ref{sec:private}), firms can coordinate their actions based on the public monitoring outcome announced by a central monitor. 

In the remainder of this section, we formalize this intuition by first presenting the folk theorem of \cite{fudenberg94} for infinitely repeated games with imperfect public monitoring in Section \ref{sec:folk-public}, \rev{and verify the applicability of this folk theorem to NPD information sharing games with monitoring given by \eqref{eq:public-mon} in Section \ref{sec:verify-public}}. We then present a numerical example to illustrate the construction of equilibrium strategies based on this folk theorem in Section \ref{sec:example-public}.

\subsection{The Folk Theorem with Imperfect Public Monitoring} \label{sec:folk-public}
We briefly present the folk theorem due to \cite{fudenberg94}; we refer the interested reader to that paper, or \cite{naghizadeh16b}, for a more detailed description. The setup is largely similar to that of Section \ref{sec:folk-private}, with one major distinction. Given the availability of a publicly observable signal, firms have the option of fully ignoring their own private histories, and choosing their actions solely based on the public history $h^{t-1}$. 
Such strategies are known as \emph{public strategies}. Whenever other firms are playing public strategies, then firm $i$ will also have a public strategy best-response; see \cite{mailath06}. 

The equilibrium concept of interest is then a \emph{perfect public equilibrium (PPE)}.} This is defined as a profile of public strategies that, starting at any time $t$ and given any public history $h^{t-1}$, form a Nash equilibrium of the game from that point on. PPEs facilitate the study of repeated games to a great extent, as they are ``recursive''. This means that when a PPE is being played, the continuation game at each time point is strategically isomorphic to the original game, and therefore the same PPE is induced in the continuation game as well.\footnote{Note that such recursive structure can not be recovered using private strategies, leading to the comparatively limited results in private monitoring games, as discussed in Section \ref{sec:private}.} Let $\mathcal{E}(\delta)$ be the set of all payoff profiles that can be attained using public strategies as PPE average payoffs when the discount factor is $\delta$. The main goal of the folk theorem is again to identify conditions under which it is possible to attain any point in the interior of the set $\mathcal{F}^*$ of feasible and strictly  individually rational payoffs as PPE payoffs.    

In order to coordinate on implementing desired payoffs, firms again need to be able to support cooperation by detecting and appropriately punishing deviations from the desired levels of information sharing by any firm. In PPEs, where strategies are public, all such punishment should occur solely based on the public signals. As a result, the public signals should be distributed so as to satisfy two conditions. The first condition, referred to as \emph{individual full rank}, gives a sufficient condition under which deviations by a single firm are statistically distinguishable. 
Formally, 

\begin{condition} 
The profile $\boldsymbol{r}$ has individual full rank for firm $i$ if given the strategies of the other firms, $\boldsymbol{r_{-i}}$, the $|R_i|\times|B|$ matrix $A_{i}(\boldsymbol{r_{-i}})$ with entires $[A_{i}(\boldsymbol{r_{-i}})]_{{r_i, \mathbf{b}}} = \pi(\mathbf{b}|r_i, \boldsymbol{r_{-i}})$ has full row rank. That is, the $|R_i|$ vectors $\{\pi(\cdot | r_i, \boldsymbol{r}_{-i})\}_{r_i\in R_i}$ are linearly independent. 
\end{condition}

The second general condition, \emph{pairwise full rank}, is a strengthening of individual full rank to pairs of firms. In essence, it ensures that deviations by firms $i$ and $j$ are distinct, as they introduce different distributions over public outcomes. Formally, 

\begin{condition}
 The profile $\boldsymbol{r}$ has pairwise full rank for firms $i$ and $j$ if the $(|R_i|+ |R_j|)\times|B|$ matrix $A_{ij}(\boldsymbol{r}):=[A_{i}(\boldsymbol{r_{-i}}); A_{j}(\boldsymbol{r_{-j}})]$ has rank $|R_i|+ |R_j|-1$.
\end{condition}

{It is worth noting that (C4) can be viewed as a stronger version of (C1), and (C5) is a stronger version of Conditions (C2) and (C3); see \cite{kandori98}.} 
Given an adequate public monitoring signal, we have the following folk theorem. 

\begin{theorem} {\sc{(The imperfect public monitoring folk theorem \cite{fudenberg94})}.}\label{thm:publicfolk} 
Assume $R$ is finite, the set of feasible payoffs $\mathcal{F}^\dagger \subset \mathbb{R}^N$ has non-empty interior, and all the pure action equilibria leading the extreme points of $\mathcal{F}^\dagger$ have pairwise full rank for all pairs of firms. If the minmax payoff profile $\underline{\mathbf{v}}=(\underline{v}_1, \ldots, \underline{v}_N)$ is inefficient, and the minmax profile leading to these payoffs 
has individual full rank for each firm $i$, then for any profile of payoffs $\mathbf{v}\in \text{int}\mathcal{F}^*$, there exists a discount factor $\underline{\delta}<1$, such that for all $\delta\in (\underline{\delta}, 1)$, $\mathbf{v} \in \mathcal{E}(\delta)$. 
\end{theorem}

\subsection{\rev{Cooperation in Information Sharing Agreements}}
\label{sec:verify-public}
{To show that this folk theorem holds in the information sharing games with the  monitoring structure given by \eqref{eq:public-mon}, we need to verify that the conditions on the informativeness of the public signal hold for \eqref{eq:public-mon}. This is verified in the following lemma; the proof is given in the appendix.  
\begin{lemma}\label{lemma:public}
The conditions (C4) and (C5) of the folk theorem in Theorem \ref{thm:publicfolk} hold with public monitoring distribution given in \eqref{eq:public-mon}. 
\end{lemma}
Lemma \ref{lemma:public} together with Theorem \ref{thm:publicfolk} establish the following.} 
\begin{proposition}
The conditions of the folk theorem in Theorem \ref{thm:publicfolk} hold with the public monitoring distribution given in \eqref{eq:public-mon}. As a result, when the firms are sufficiently patient, it is possible for them to nearly efficiently cooperate on full information disclosure through repeated interactions. 
\end{proposition}
Therefore, when firms are sufficiently patient, they can sustain cooperation on full security information sharing in a repeated setting, by making their disclosure decisions based only on the imperfect, publicly announced observations of the monitor about their past actions.

\subsection{Constructing Public Strategies: An Example}\label{sec:example-public}
{So far, in both Sections \ref{sec:verify-private} and \ref{sec:verify-public}, we only verified the possibility of constructing equilibrium strategies leading to a desired payoff profile in infinitely repeated interactions, without explicitly specifying firms' strategies at each stage. In this section, we present a numerical example to illustrate a \emph{process} through which equilibrium public strategies can be constructed.} To simplify the illustration, we consider a two player prisoner's dilemma game with payoff matrix given by Table \ref{t:pdgame}. 

\begin{table}
\centering
\caption{Firms' payoffs in a two-person prisoner's dilemma game.}\label{t:pdgame}
{\begin{tabular}{c|c|c|}
\multicolumn{1}{r}{}
 &  \multicolumn{1}{c}{C}
 & \multicolumn{1}{c}{D} \\
  \cline{2-3}
C & $G-L$, $G-L$ & $-L$, $G$ \\
  \cline{2-3}
D & $G$, $-L$ & 0, 0 \\
  \cline{2-3}
\end{tabular}
}
\end{table}

We first present an overview of the idea behind constructing the equilibrium strategies. The utility of firms at each step of the game can be decomposed into their current payoff, plus the continuation payoff, i.e., the expected payoff for the remainder of the game depending on the observed public monitoring output. Therefore, to achieve an average payoff profile $\mathbf{v}$ as equilibrium in the repeated game, the action profile and the continuation payoffs should be selected so as to maximize firms' expected payoff. 

Formally, we say $\mathbf{v}$ is decomposed by $\mathbf{r}$ on a set $W$ using a mapping $\gamma: B\rightarrow W$ if: 
\begin{align}  \label{eq:decompose}
v_i &=  (1-\delta)u_i(\mathbf{r}) + \delta E[\gamma_i(b)|\mathbf{r}]\\ 
     &\geq (1-\delta)u_i(r_i', \mathbf{r}_{-i}) + \delta E[\gamma_i(b)|r_i', \mathbf{r}_{-i}], ~\forall r_i'\in R_i~, \forall i.\notag
\end{align}
Here, the mapping $\gamma$ determines firms' continuation payoffs (selected from a set $W$) following each signal $b\in B$. The goal is thus to set $W=\mathcal{E}(\delta)$ (the set of PPE payoffs), and find appropriate actions $\mathbf{r}$ and mappings $\gamma$ decomposing (i.e., satisfying \eqref{eq:decompose} for) payoff profiles $\mathbf{v}\in \mathcal{E}(\delta)$. We can then conclude that any payoff profile $\mathbf{v}$ for which the above decomposition is possible will be attainable as a PPE average payoff, as we can recursively decompose the selected continuation payoffs on $\mathcal{E}(\delta)$ as well. This procedure thus characterizes the set of payoffs that can be attained using public strategies. 

However, the set of decomposable payoffs on arbitrary sets $W$ is in general hard to characterize; let's instead consider the simpler decomposition on half-spaces $H(\lambda, \lambda \cdot \mathbf{{v}}):=\{\mathbf{v'}\in \mathbb{R}^N: \lambda\cdot\mathbf{v'}\leq \lambda\cdot\mathbf{{v}}\}$. With $W=H(\lambda, \lambda \cdot \mathbf{{v}})$, \eqref{eq:decompose} can be re-written as:
\begin{align}
v_i ~ = & ~~ u_i(\mathbf{r}) + E[\bar{\gamma}_i(b)|\mathbf{r}] 
    ~~ \geq ~~ u_i(r_i', \mathbf{r}_{-i}) + E[\bar{\gamma}_i(b)|r_i', \mathbf{r}_{-i}],\notag\\
    & ~\forall r_i'\in R_i~, \forall i~,
     \text{ and, } ~~ \lambda \cdot \bar{\gamma}(b) \leq 0,~ \forall b\in B~,
     \label{eq:k-star}
\end{align}
where $\bar{\gamma}: B\rightarrow \mathbb{R}^N$, and $\bar{\gamma}_i(b)=\frac{\delta}{1-\delta}(\gamma_i(b)-v_i)$. We refer to $\bar{\gamma}$ as the normalized continuation payoffs. 

It can be shown (see \cite{mailath06}) that characterizing the set of attainable PPE payoffs $\mathcal{E}(\delta)$ is equivalent to finding the maximum average payoffs that can be decomposed on half-spaces using different actions $\mathbf{r}$ and in various directions $\lambda$. We therefore first find the maximum average payoffs $\mathbf{v}$ enforceable on half-spaces (i.e, satisfying \eqref{eq:k-star}, and with $\lambda\cdot \bar{\gamma}(b)=0$ whenever possible), for each action profile $\mathbf{r}$ and direction $\lambda$. We will then select the best action $\mathbf{r}$ for each direction, and finally take the intersection over all possible directions $\lambda$ to characterize $\mathcal{E}(\delta)$.\footnote{Define $k^*(\lambda; \mathbf{r}) := \lambda\cdot \mathbf{\bar{v}}$, where  $\mathbf{\bar{v}}$ is the maximum payoff profile satisfying \eqref{eq:k-star}. It can be shown that $k^*(\lambda; \mathbf{r}) \leq \lambda\cdot u(\mathbf{r})$, and so the maximum is attained when $\mathbf{r}$ is \emph{orthogonally enforced} (whenever possible), i.e., $\lambda\cdot \bar{\gamma}(b)=0$ in \eqref{eq:k-star}. Let $k^*(\lambda) = \sup_{\mathbf{r}} k^*(\lambda; \mathbf{r})$. Intuitively, $k^*(\lambda)$ is a bound on the average payoff for firms for which the incentive constraints are satisfied. Let $H^*(\lambda):=H(\lambda, k^*(\lambda))$ be the corresponding \emph{maximal half-space}. Then, that the set of PPE payoffs is contained in the intersection of these maximal half-spaces, i.e., $\mathcal{E}(\delta)\subseteq\cap_{\lambda}H^*(\lambda):=\mathcal{M}$, and that the reverse is also true for sufficiently large $\delta$, i.e, $\lim_{\delta\rightarrow 1} \mathcal{E}(\delta)=\mathcal{M}$. We refer the interested reader to \cite{mailath06} for more details.}

To find the average payoffs decomposable on half-spaces for the prisoner's dilemma game in Table \ref{t:pdgame}, we first consider profile $\mathbf{r}=(1,1)$,\footnote{Note that decomposing using $(0,0)$ is not considered as it leads to the maximal half-space $\mathbb{R}^2$. It thus provides no information on the set of attainable payoffs as we already know that $\mathcal{E}(\delta)\subseteq \mathbb{R}^2$.} 
and an arbitrary direction $\lambda=(\lambda_1, \lambda_2)$. Setting $\lambda\cdot \bar{\gamma}(b)=0$, \eqref{eq:k-star} reduces to:
{\begin{small}
\begin{align*}
G-L &= G-L + (\epsilon^2 \bar{\gamma}_1(0,0) + \epsilon (1-\epsilon) \bar{\gamma}_1(0,1) +\notag\\
& \qquad (1-\epsilon) \epsilon \bar{\gamma}_1(1,0) + (1-\epsilon)^2 \bar{\gamma}_1(1,1)) \notag\\
 &\geq ~~ G ~ + (\epsilon \alpha \bar{\gamma}_1(0,0) + \alpha (1-\epsilon) \bar{\gamma}_1(0,1) + \notag\\
& \qquad (1-\alpha) \epsilon \bar{\gamma}_1(1,0) + (1-\epsilon) (1-\alpha) \bar{\gamma}_1(1,1)) \notag\\
&\text{ and~,}\notag\\
G-L &= G-L  + (\epsilon^2 \bar{\gamma}_2(0,0) + \epsilon (1-\epsilon) \bar{\gamma}_2(0,1) + \notag\\
& \qquad (1-\epsilon) \epsilon \bar{\gamma}_2(1,0) + (1-\epsilon)^2 \bar{\gamma}_2(1,1)) \notag\\
 &\geq ~~ G ~ + (\epsilon \alpha \bar{\gamma}_2(0,0) + \epsilon (1-\alpha) \bar{\gamma}_2(0,1) + \notag\\
& \qquad (1-\epsilon) \alpha \bar{\gamma}_2(1,0) + (1-\epsilon) (1-\alpha) \bar{\gamma}_2(1,1)) \notag\\
 &\text{ and~,}\notag\\
 & \lambda_1 \bar{\gamma}_1(b) + \lambda_2 \bar{\gamma}_2(b) = 0, ~~ \forall b\in B~.
\end{align*}
\end{small}}
Substituting for $\bar{\gamma}_2(b)$ using the last equation, and writing the inequalities as equalities, finding the normalized continuation payoffs is equivalent to solving: 
\[\begin{psmallmatrix}
\cr
                  & \epsilon^2 & \epsilon(1-\epsilon) & (1-\epsilon)\epsilon & (1-\epsilon)^2\cr
		 & \alpha\epsilon & \alpha(1-\epsilon) & (1-\alpha)\epsilon & (1-\alpha)(1-\epsilon) \cr
                  & \epsilon^2 & \epsilon(1-\epsilon) & (1-\epsilon)\epsilon & (1-\epsilon)^2 \cr
                  & \alpha\epsilon & (1-\alpha)\epsilon & \alpha(1-\epsilon) & (1-\alpha)(1-\epsilon) \cr
\end{psmallmatrix}
\begin{psmallmatrix}
\cr & \bar{\gamma}_1(0,0)\cr &\bar{\gamma}_1(0,1)\cr &\bar{\gamma}_1(1,0) \cr & \bar{\gamma}_1(1,1)\cr
\end{psmallmatrix} = 
\begin{psmallmatrix}
\cr & 0 \cr & -L \cr & 0 \cr & L\frac{\lambda_2}{\lambda_1}\cr
\end{psmallmatrix}\]
The first and third rows represent the same equations (corresponding to the equilibrium outcome). Removing the third row and performing row-reduction on the remaining matrix, the continuation payoffs should satisfy the following set of equations: 
\begin{align*}
\epsilon \bar{\gamma}_1(0,0) + (1-\epsilon) \bar{\gamma}_1(0,1) &= \frac{-L}{\alpha \kappa} \frac{1-\epsilon}{\epsilon}\notag\\
\epsilon \bar{\gamma}_1(1,0) + (1-\epsilon) \bar{\gamma}_1(1,1) &= \frac{L}{\alpha \kappa} \notag\\
- \bar{\gamma}_1(0,1) + \bar{\gamma}_1(1,0) &= \frac{L}{\epsilon \alpha \kappa} (\frac{\lambda_2}{\lambda_1} + 1)~,
\end{align*}
where $\kappa := \frac{1-\epsilon}{\epsilon} - \frac{1-\alpha}{\alpha}>0$. The above is an underdetermined system, and thus has infinitely many solutions depending on the designer's choice of continuation payoffs. We construct and interpret one such possibility. 

Let's set $\bar{\gamma}_1(1,1)=0$, implying $\bar{\gamma}_2(1,1)=0$ as well. This means if the signal indicates that both firms are cooperating with high probability, there is no need for punishments, so both firms expect their continuation payoff to remain unchanged (i.e., equal to their current payoff). Given this choice, we can solve for the remaining normalized continuation payoffs, illustrated in Table \ref{t:continuations}. 

\begin{table}
\centering
\caption{An example of normalized continuation payoff choices.}\label{t:continuations}{
\begin{tabular}{c|c|c|}
\multicolumn{1}{r}{}
 &  \multicolumn{1}{c}{$\bar{\gamma}_1(b)$}
 & \multicolumn{1}{c}{$\bar{\gamma}_2(b)$} \\
  \cline{2-3}
b=(0,0) & $\frac{L}{\epsilon \alpha \kappa} \frac{1-\epsilon}{\epsilon} (\frac{\lambda_2}{\lambda_1}-1)$ & $ \frac{L}{\epsilon \alpha \kappa} \frac{1-\epsilon}{\epsilon} (\frac{\lambda_1}{\lambda_2}-1)$ \\
  \cline{2-3}
b=(0,1) & $-\frac{\lambda_2}{\lambda_1}\frac{L}{\epsilon \alpha \kappa}$ & $\frac{L}{\epsilon \alpha \kappa}$ \\
  \cline{2-3}
b=(1,0) & $\frac{L}{\epsilon \alpha \kappa}$ & $- \frac{\lambda_1}{\lambda_2}\frac{L}{\epsilon \alpha \kappa}$ \\
  \cline{2-3}
b=(1,1) & 0 & 0 \\
  \cline{2-3}
\end{tabular}
}
\end{table}

{\bf Interpretation of continuation payoffs.} These normalized continuation payoffs can be intuitively interpreted as follows. Fix a direction with $\lambda_1, \lambda_2>0$. 
Then, given a signal $b=(1,0)$, which is more likely under a deviation by firm 2, firm 1 expects a higher continuation payoff ($\bar{\gamma}_1(1,0)>0$), while the suspect deviator expects a lower one ($\bar{\gamma}_2(1,0)<0$).\footnote{It is worth emphasizing that due to the equilibrium construction, firms are both playing $r_i=1$; nevertheless, punishments on the equilibrium path happen due to the imperfection of monitoring.} A similar intuition applies to the continuations under the signal $(0,1)$. On the other hand, with $b=(0,0)$, either firm 1 or 2 will be punished, depending on the direction $\lambda$. Specifically, for a direction $\lambda_1=\lambda_2$, neither firm expects a change in her continuation payoff. Note that with $\lambda_1=\lambda_2$, the change in continuation payoffs between the outcomes $(0,1)$ and $(1,0)$, as well as among firms in either outcome, are also of equal size. Note also that both firms are never punished simultaneously under any outcome, so as to maintain a high average payoff. 

\begin{figure}
\centering
  \centering
  \includegraphics[width=0.64\linewidth]{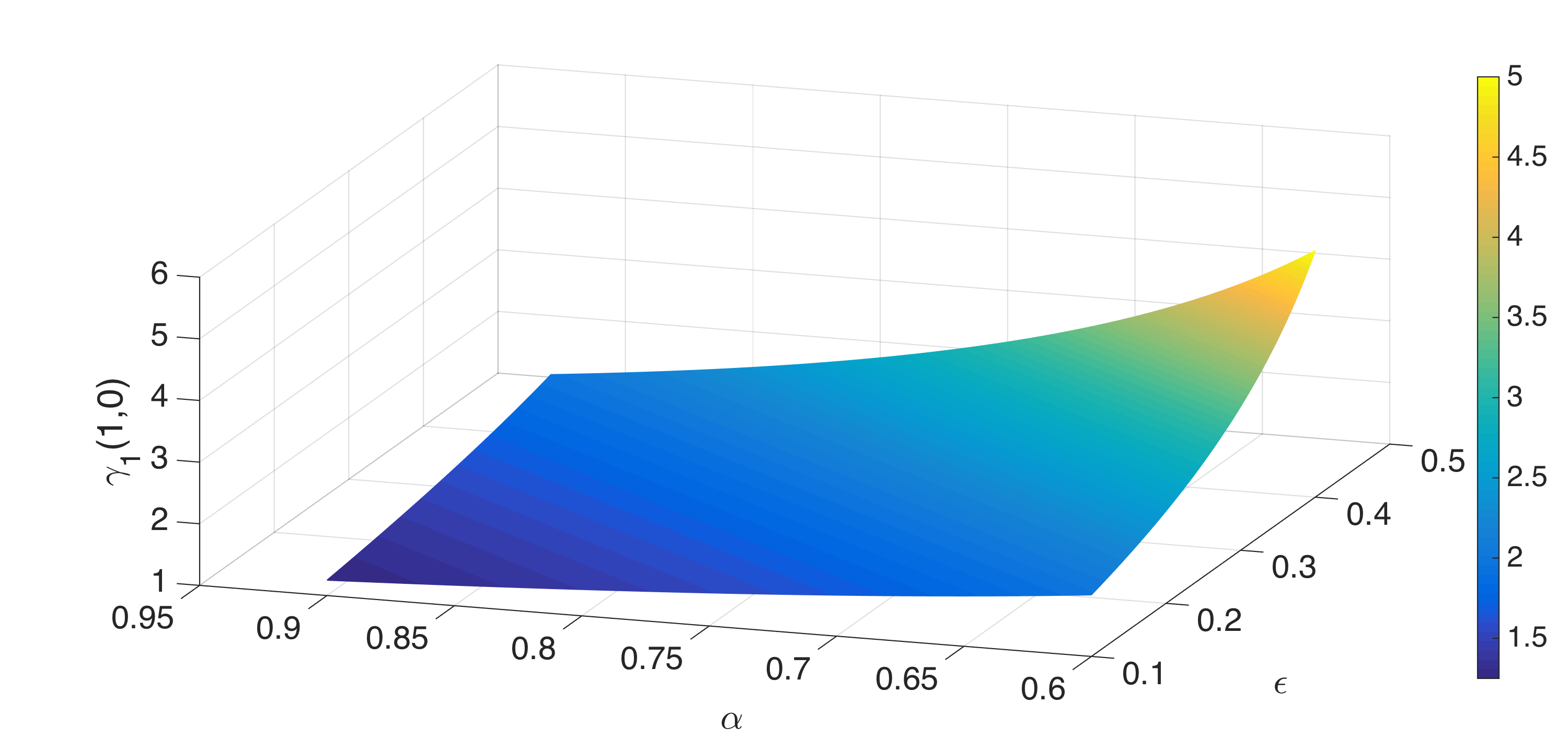}
  \caption{$\bar{\gamma}_1(1,0)$}
  \label{fig:x13}
\end{figure}

{\bf Effects of monitoring accuracy.} Finally, it is worth noting the effect of the monitoring accuracy, $\alpha$ and $\epsilon$, on the normalized continuation payoffs. Consider direction $\lambda_1=\lambda_2=1$, and fix $L=1$. First, note that $\epsilon\alpha\kappa=\alpha(1-\epsilon) - \epsilon(1-\alpha)$ is increasing in $\alpha$ and decreasing in $\epsilon$. This is illustrated in Fig. \ref{fig:x13}, which shows the dependence of $\bar{\gamma}_1(1,0)$ on the monitoring parameters. As a result, as the monitoring technology becomes more accurate, i.e., $\alpha$ increases and/or $\epsilon$ decreases, the size of the normalized continuation payoffs for firms, when $(1,0)$ or $(0,1)$ is observed, becomes smaller. This is because, as monitoring becomes accurate, signals indicating deviations (despite equilibrium being played) happen only due to decreasing monitoring errors (rather than actual deviations), and therefore the required continuation punishments/rewards for off-equilibrium paths can become less severe, while still maintaining a high average payoff.

We conclude that in general, using the described procedure, we can decompose payoff profiles in the half-spaces $H(\lambda, k^*(\lambda, (1,1)))$, where $k^*(\lambda, (1,1))=\lambda\cdot u(1,1) = (G-L)(\lambda_2 + \lambda_1)$,  using the action profile $\mathbf{r}=(1,1)$ and continuation payoffs determined as above. Using a similar procedure, the corresponding half-spaces for the remaining action profiles will have $k^*(\lambda, (0,1)) = G\lambda_1 - L\lambda_2$ and $k^*(\lambda, (1,0)) = G\lambda_2 - L\lambda_1$. 

We next choose, for a given direction $\lambda$, the action for which the corresponding half-spaces covers a larger set of average payoffs, i.e, $k^*(\lambda) = \max_{\mathbf{r}}\{G\lambda_2 - L\lambda_1, G\lambda_1 - L\lambda_2, (G-L)(\lambda_1+\lambda_2)\}$; which leads to:
\begin{align*}
k^*(\lambda) = \begin{cases} 
G\lambda_2 - L\lambda_1 & \lambda_2 \geq \frac{G}{L} \lambda_1\\
(G-L)(\lambda_1+\lambda_2) & \frac{L}{G} \lambda_1 \leq \lambda_2 \leq \frac{G}{L} \lambda_1\\
G\lambda_1 - L\lambda_2 & \lambda_1 \geq \frac{G}{L} \lambda_2
\end{cases}
\end{align*} 
Finally, it is straightforward to show that the intersection of half-spaces $H(\lambda, k^*(\lambda))$, as $\lambda$ ranges over $\mathbb{R}^2$, is equivalent to the set of feasible and strictly individually rational payoffs of the two-person prisoner's dilemma game of Table \ref{t:pdgame}. That is, it is possible to find an action profile $\mathbf{r}$ and the corresponding continuation payoff mapping $\gamma$ (constructed as described above), so as to incentivize any feasible strictly individually rational payoff profile.

\section{Discussion} \label{sec:disc}
\subsection{Practical and Policy Implications}\label{sec:implications} {
We now discuss some practical implications of our findings. 
We first note that {``trust'', which in this context refers to anticipating reciprocal information sharing and collaboration,} is often anecdotally cited as a factor impacting the formation and operation of ISACs, see e.g., \cite{scmagazine,dhsmeeting,rcisc}. %
The NIST Guide to Cyber Threat Information Sharing \cite{nist} states that ``peer-to-peer trust is based on the belief that peers support a common mission, respect the established sharing rules, and demonstrate a willingness to participate in reciprocal sharing'', and ``ongoing communication, through regular in-person meetings, phone calls, or social media can help accelerate the process of building trust''. Our work formalizes these notions of private monitoring and communication, and proposes history-dependent incentive mechanisms for sustaining the {trust in future cooperative behavior} among participants. 

Further, there is a close parallel between our models of public and private monitoring, and the information sharing approaches identified by the MITRE Corporation \cite{mitre}. Specifically, \cite{mitre} classifies information sharing structures into Hub-and-Spoke, where participants share information with a central entity who later disseminates the information, and Post-to-All models, in which participants share information with all other participants directly. Our proposed public and private monitoring components apply to the Hub-and-Spoke and Post-to-All models, respectively. 

We next discuss the availability of monitoring technologies. Public monitoring of firms' security scores, based on externally observable security indicators, is emerging as a quantitative and rigorous field.  A number of products introduced in the market in the last few years attempt to provide (near) global security ratings for enterprises. Examples include the FICO Enterprise Security Scores (ESS), Security Scorecard, and BitSight Technologies.\footnote{See \url{http://www.fico.com/en/products/fico-enterprise-security-scoring}, \url{http://securityscorecard.com}, and \url{http://www.bitsighttech.com}.} While these products do not monitor firms' information disclosure decisions, they provide a quantitative assessment of a firm's security risks. For instance, the data breach prediction technology underlying the FICO ESS product, as reported in \cite{liu15}, quantifies the prediction accuracy with both true positive and false positive probabilities.
In a simplified case, a high risk assessment paired with lack of any problems shared by that organization may lead to the belief that the corresponding organization is being untruthful, with the same true positive and false positive probabilities associated with the assessment tool. The outcome of such assessments can therefore be used as a form of imperfect public monitoring. 

In addition, instances of private monitoring are likely implicitly and informally done within organizations engaged in an information sharing agreement with one another. In the case of ISACs, despite a prior assumption that firms will adhere to the terms of the agreement, as parties interact over time, mutual trust may strengthen or deteriorate, depending on what transpires. For instance, distrust may be triggered if an undisclosed incident is unveiled due to other business partnerships, or if representatives from an organization are consistently absent from an ISAC's scheduled meetings. This could further lead other participants to gather their own information on the distrusted entities, which constitutes a form of private monitoring.  
An ISAC participant can then take the publicly assessed risk of another participant (using one of the aforementioned rating products), and combine it with the information it receives directly from that participant through their interactions in and out of the agreement, to form a belief about whether the latter is being truthful.  Our results state that by providing a platform to communicate such beliefs, firms can coordinate on improving disclosure levels in the agreement. 

\rev{As an alternative to these intra-firm (privately managed) ISACs, our findings have further established the feasibility of achieving similar inter-temporal incentives through centralized assessments by a common monitor. From a policy perspective, this presents authorities with a potential alternative for collecting and disseminating security information from incidents that do not fall under the umbrella of mandatory disclosure laws. As long as information gains outweigh individual disclosure costs, there can be incentives for firms to  voluntarily opt into these centers, and comply with voluntary disclosure requirements (as monitored by the organizing authority) to benefit from continued membership.}

\rev{Ultimately, the feasibility of sustaining long-run cooperation in information sharing agreements using our proposed inter-temporal incentives will depend on the firms' ``patience'' (i.e. value placed on future information/interactions). Our results state that for inter-temporal incentives to incentivize cooperation, firms have to be sufficiently patient. On the other hand, improving the precision of monitoring can increase the space of outcomes that can be supported as equilibria in repeated interactions \cite{mailath06}. Together, this means that in practice, monitoring accuracy may need to be improved in order to sustain cooperation among short-sighted firms. Firms' incentives for cooperation will further depend on their evaluation of gains from attained information and losses from breach disclosure. Such assessments can be done both internally and externally, e.g. by analyzing market reactions to breach disclosures \cite{campbell03,cavusoglu04}. The availability and accuracy of these estimations can further shape (short-sighted) firms' participation incentives.}

We close this section by discussing the \emph{necessity} for communication in sustaining cooperation, from a technical viewpoint. The folk theorem of \cite{kandori98} establishes the \emph{possibility} of sustaining cooperation under private monitoring and communication. 
\cite{awaya14} further shows that in the prisoner's dilemma game with fixed discount rate, communication is in fact \emph{necessary} for sustaining cooperation. Our work is similarly motivated by the (technical and intuitive) need for communication in order to sustain cooperation in information sharing agreements.}

\rev{
\subsection{Extensions and Limitations of Our Model}\label{sec:limitations}
The findings of this paper will continue to hold under several extensions to our proposed model of security information sharing in Section \ref{sec:model}. In particular, the folk theorem will continue to hold for extensions to both firms' disclosure decisions $r_i$ and monitoring beliefs $b_{ij}$ from binary decisions to discrete finite sets. Furthermore, while we have proposed the framework of N-player prisoner's dilemma games as a way to capture conflicts in gains and losses from information sharing, the folk theorems of Theorems \ref{thm:private} and \ref{thm:publicfolk} are not limited to the utility functions of the form \eqref{eq:utilities}. In particular, our findings can be extended to payoffs with asymmetric disclosure losses and asymmetric gains of information from different firm's disclosed information. 

In terms of the monitoring structure, \ref{eq:private-mon} assumes that firms' monitoring technologies have homogenous accuracy. For heterogenous accuracies, as well as other monitoring functions, we will need to verify that the collective belief gathered through the communication platform can satisfy the ``informativeness'' conditions of Theorem \ref{thm:private}. If belief accuracies are known, an aggregate belief profile may be constructed accordingly; for instance, it may be sufficient to rely on the most informative signal about each firm as the representative belief. Our framework does not however capture firms' incentives for falsifying their beliefs about others (due to e.g. rivalry). Determining optimal belief aggregation in such general settings is an interesting extension. 

Lastly, as mentioned in Section \ref{sec:model}, our model focuses on firms' breach reporting decisions. Capturing the interplay between firms' reporting decisions and security investment decisions (including the resulting positive/negative externalities), and allowing for the spread of mis-information as a potential action, remain interesting directions of future work. 
}

\section{Related Work} \label{sec:related}

A number of research papers have analyzed the implications of {laws for breach notification to authorities, as well as information sharing agreements among firms.} {We refer the interested reader to a recent survey by Laube and B{\"o}hme \cite{laube2017strategic} for a systematic review of theoretical work as well as empirical work studying the strategic aspects of security information sharing. Below, we summarize the work most closely related to the current paper.} 

{Breach reporting to \emph{an authority} has been studied by} Laube and B{\"o}hme \cite{laube16}. 
They study the effectiveness of mandatory breach reporting, and show that enforcing breach disclosure to an authority (using audits and sanctions) is effective in increasing social welfare only under certain conditions, including high interdependence among firms and low disclosure costs. {Further, the work of Ogut et al. \cite{ogut05}, which studies firms' investments in IT security and cyber insurance, considers information sharing with authorities as a potential mechanism to improve firms' incentives. They show that if the availability of shared information can reduce either attack probabilities or firms' interdependency, it will benefit social welfare by inducing firms to improve investments in self-protection and cyber-insurance.} 

{Several studies have analyzed game-theoretic models of information sharing \emph{among firms}.} 
Gordon et al. \cite{gordon03} show that, if security information from a partner firm is a \emph{substitute} to a firm's own security expenditures, then (mandatory) information sharing laws reduce expenditure in security measures, but can nevertheless increase social welfare. However, firms will not voluntarily comply with sharing agreements, requiring additional economic incentives to be in place (e.g., a charge on a member of the ISAC for losses on the other member). 
Gal-Or and Ghose \cite{gal05} on the other hand allow information sharing to be a \emph{complement} to the firm's own security expenditures, as it may increase consumer confidence in a firm that is believed to take steps towards securing its system. Using this model, the authors show that when the positive demand effects of information sharing are high enough, added expenditure and/or sharing by one firm can incentivize the other firm to also increase its expenditure and/or sharing levels. {Hausken \cite{Hausken07} introduces an external attacker as a decision maker to a similar model, and emphasizes firms' levels of interdependency as a key factor in determining information sharing. Liu et al. \cite{liu11} on the other hand introduce the possibility of firms having information assets that are complementary or substitutable to another firm's assets. They show that while complementary assets will lead to voluntary sharing of information, substitutable assets result in a prisoner's dilemma scenario, in which sharing does not occur despite its social benefits.}

In this work, 
we assume disclosure costs are higher than potential demand-side benefits, therefore  predicting a lack of voluntary information sharing at equilibrium similar to several of the works discussed above. Our proposed approach of considering the effects of repeated interactions as an incentive solution is however different from those proposed in aforementioned literature, as they consider one-shot games.

{Our conclusions are also in line with the empirical study of information sharing agreements in \cite{mermoud2018incentives}. In particular, Marmoud et al. \cite{mermoud2018incentives} conduct an empirical study of information sharing agreements by surveying participants in a Swiss government-organized ISAC. Their analysis shows that both the intensity and frequency of information sharing are influenced by the firms' expectation of (social) reciprocity. These findings are in line with the premise of our proposed mechanisms, which formalize the use of trust and reciprocity to further collaboration in information sharing agreements.} 

\rev{More broadly, our work falls within the literature on incentivizing information sharing in repeated games. Similar ideas have been explored in other contexts. For instance, in the context of service delivery, Heegaard et al. \cite{heegaard16} study the effects of information sharing between 2 operators on the network users' QoS, and show that inter-temporal incentives can be provided given the applicability of a folk theorem with \emph{perfect} public monitoring. In the context of distributed multi-agent optimization, Yu et al. \cite{yu2015} propose the use of reputation scores by agents to decide future information exchange based on others' (perfectly observed) past behavior. In the context of spectrum sharing, Teng et al. \cite{teng17} study incentives for operators to communicate their private information about their \emph{own} traffic intensities; however, monitoring of deviations from equilibrium strategies is common between operators. Our prior work \cite{naghizadeh16b} studies a 2 firm model of security information sharing games with public monitoring. While our use of inter-temporal incentives is similar in nature to \cite{naghizadeh16b,heegaard16,yu2015,teng17}, this paper takes into account the inevitable {imperfectness}, as well as potentially private nature, of monitoring of firms' security information sharing decisions. We therefore focus on the applicability of folk theorem for repeated games with \emph{imperfect} public and \emph{private} monitoring \cite{mailath06}. 
}

\section{Conclusion} \label{sec:conclusion}

We modeled information sharing agreements among firms as an N-person prisoner's dilemma game equipped with a simple monitoring structure. We proposed a repeated-game approach to this problem, and discussed the role of monitoring (private vs. public) in building inter-temporal incentives that can lead to firms' cooperation on full disclosure. Specifically, we showed that firms can fully cooperate in the long run when provided with a platform to communicate their privately observed beliefs on each others' adherence to the agreement. A similar result can be attained if firms coordinate their sharing decisions based on reports by a central monitor.

An important requirement for the folk theorem, and consequently the design of inter-temporal incentives, is to ensure that firms are sufficiently patient (i.e., they place significant value on their future interactions), as characterized by having discount factors higher than $\underline{\delta}$. Despite the fact that the proposed binary monitoring structures in \eqref{eq:private-mon} and \eqref{eq:public-mon} are informative enough for the folk theorem to hold, their accuracy, $(\alpha, \epsilon)$, will impact the requirement on firms' patience, $\underline{\delta}$. Characterizing the dependence of $\underline{\delta}$ on  $(\alpha, \epsilon)$ is a main direction of future work. 

Another possible direction is to consider the design of inter-temporal incentives when both types of public and private monitoring are available. It is indeed still possible to have firms coordinate based on the public monitoring system's report alone (i.e., use public strategies); nevertheless, it may also be possible to employ \emph{private strategies}, in which firms use both their own observations, as well as the public signal. Private strategies may lead to higher payoffs than those attainable through public strategies alone \cite[Chapter 10]{mailath06}, thus making their study of interest to either lower the required discount factor, or when the monitoring signals are not informative enough for a public monitoring folk theorem to hold. 

Finally, we have assumed that the monitoring, as well as its accuracy, are fixed and available to firms at no additional cost. Analyzing the effects of costly monitoring on firms' incentives is another direction of future work.

\appendix

\subsection*{Proof of Lemma \ref{lemma:private}}
Once firms' private beliefs are truthfully reported, we have access to $N-1$ independent realizations of the distribution in \eqref{eq:private-mon}. That is, as the signal distributions of the non-deviators are iid, to test the conditions of the folk theorem, it is sufficient to randomly choose one of the available cross-observations about possible deviator(s), from the firms other than the deviator(s). The collection of samples selected from \eqref{eq:private-mon} can be in turn viewed as a sample of the distribution \eqref{eq:public-mon}. Thus, equivalently, we can verify the conditions of the folk theorem in Theorem \ref{thm:private} on a joint distribution of private signals given by \eqref{eq:public-mon}. 

We first verify Condition (C1) that, for any firm $i$, the minmax profile of the repeated information sharing game leads to distinguishable distributions on other firms' private beliefs. The minmax action profile for some firm $i$, $\mathbf{\hat r}^i$, is all firms concealing their information, i.e., $\mathbf{\hat r}^i = \mathbf{0}$. Consider deviations by firm $1$ (the same argument holds for other firms). Then $p_{-i}(\mathbf{\hat r}^i)$ is given by:
\begin{scriptsize}
\begin{align*}
\bordermatrix{~~\mathbf{b} =  & (0,0,\ldots,0) & (1,0,\ldots,0) & \ldots 
& (1, 1, \ldots, 1) \cr
                  r_1=0 &\alpha^N & (1-\alpha)\alpha^{N-1}& \ldots 
                  & (1-\alpha)^{N} \cr
                  r_1=1 &\epsilon\alpha^{N-1} & (1-\epsilon)\alpha^{N-1} & \ldots 
                  & (1-\epsilon)(1-\alpha)^{N-1} \cr}
\end{align*}
\end{scriptsize}
%
The rows of the above matrix are linearly independent (given $\alpha\neq\epsilon$), and hence the minmax profiles satisfy condition (C1). 

We next verify that the joint distribution of signals satisfies (C2) and (C3), at all pure strategy action profiles. We do so for a profile of actions $\mathbf{r}_k:=(1,1,\ldots,1, 0,0, \ldots, 0)$, in which the first $k$ firms disclose, and the remainder $N-k$ conceal; other profiles can be checked similarly. Consider two candidate deviator firms $i=1$ and $j=N$. 

We need to find $Q_{1N}(\mathbf{r}_k), p_{-\{1,N\}}(\mathbf{r}_k), Q_{N1}(\mathbf{r}_k)$, and $p_{-\{N,1\}}(\mathbf{r}_k)$ to verify the conditions (C2) and (C3). Take $Q_{1N}(\mathbf{r}_k)$ as an instance: this is the joint distribution of private beliefs given by \eqref{eq:public-mon} over all profiles $\mathbf{b}$, when firm $i$ plays action $r_1=0$ instead of $r_1=1$. When $(0,0,\ldots,0)$, for example, this will be given by $\alpha\cdot\epsilon^{k-1}\cdot\alpha^{N-k-1}\cdot\alpha $, that is, correctly observing the actions of firm 1 and firms $k+1$ through $N$, and incorrectly believing deviations from firms $2$ through $k$. 

Using a similar procedure, we construct the following matrix. The columns corresponding to profiles of beliefs $\mathbf{b}$. Note that $\mathbf{b}$ has $2^N$ possible outcomes; we view each profile as a binary string and order the columns of the following matrix are according to the decimal value of these strings. 
The first two rows correspond to $Q_{1N}(\mathbf{r}_k)$ and $p_{-\{1,N\}}(\mathbf{r}_k)$, and the last two rows correspond to $p_{-\{N,1\}}(\mathbf{r}_k)$ and $Q_{N1}(\mathbf{r}_k)$, respectively. 
\begin{scriptsize}
\[
\begin{blockarray}{cc}
		(0,0,\ldots,0) & (1,0,\ldots,0) \\
		\hspace{0.01in}\\
\begin{block}{[cc}
                   \alpha\cdot\epsilon^{k-1}\cdot\alpha^{N-k-1}\cdot\alpha 
                  &   (1-\alpha)\cdot\epsilon^{k-1}\cdot\alpha^{N-k-1}\cdot\alpha 
                   \\
                  \epsilon\cdot\epsilon^{k-1}\cdot\alpha^{N-k-1}\cdot\alpha
                  & (1-\epsilon)\cdot\epsilon^{k-1}\cdot\alpha^{N-k-1}\cdot\alpha  
		  \\
                  \epsilon\cdot\epsilon^{k-1}\cdot\alpha^{N-k-1}\cdot\alpha
                  & (1-\epsilon)\cdot\epsilon^{k-1}\cdot\alpha^{N-k-1}\cdot\alpha 
		  \\
                   \epsilon\cdot\epsilon^{k-1}\cdot\alpha^{N-k-1}\cdot\epsilon
                  & (1-\epsilon)\cdot\epsilon^{k-1}\cdot\alpha^{N-k-1}\cdot\epsilon  
		  \\
\end{block}
		  \hspace{0.1in}\\
		  \qquad \ldots 
		  & (1,1, \ldots, 1)\\
		  \hspace{0.01in}\\
\begin{block}{cc]}
                  \qquad \ldots  
                  & (1-\alpha)\cdot(1-\epsilon)^{k-1}\cdot(1-\alpha)^{N-k}\\
                  \qquad \ldots 
                  & (1-\epsilon)^{k}\cdot(1-\alpha)^{N-k}\\
                  \qquad \ldots  
                  & (1-\epsilon)^{k}\cdot(1-\alpha)^{N-k}\\
                  \qquad \ldots 
                  & (1-\epsilon)^{k}\cdot(1-\alpha)^{N-k-1}\cdot(1-\epsilon) \\
\end{block}
\end{blockarray}
\]
\end{scriptsize}
where the rows correspond to $r_1=0, r_1=1, r_N=0$, and $r_N=1$, respectively.  
Note that the rows corresponding to $r_1=1$ and $r_N=0$ are the same: indeed when both firms follow the prescribed strategy, the distribution of the signals is consistent. 
It is straightforward to verify that the above has row rank 3; i.e., removing the common row, the three remaining rows are linearly independent. 
Note that conditions (C2) and (C3) require independence in the convex combinations of the signals; this is implied by the linear independence of the signals as verified based on the matrix above. 
As a result, conditions (C2) and (C3) are satisfied for firms $i=1$ and $j=N$. A similar procedure follows for other pairs of firms $i,j$ and the remaining pure action profiles, proving the lemma.

{
\subsection*{Proof of Lemma \ref{lemma:public}}

We first verify condition (C4), showing that the minmax profile of the repeated information sharing game has individual full rank for any firm $i$. The minmax action profile for firm $i$, $\mathbf{r}^i$, is all firms concealing their information, i.e., $r^i_j=0, \forall j$. We again consider deviations by firm 1 without loss of generality. Then, $A_1(\mathbf{r}^1)$, where each column corresponds to one outcome $\mathbf{b}$ of public monitoring, is given by:
\begin{scriptsize}
\begin{align*}
\bordermatrix{~~\mathbf{b} =  & (0,0,\ldots,0) & (1,0,\ldots,0) & \ldots 
& (1, 1, \ldots, 1) \cr
                  r_1=0 &\alpha^N & (1-\alpha)\alpha^{N-1}& \ldots 
                  & (1-\alpha)^{N} \cr
                  r_1=1 &\epsilon\alpha^{N-1} & (1-\epsilon)\alpha^{N-1} & \ldots 
                  & (1-\epsilon)(1-\alpha)^{N-1} \cr}
\end{align*}
\end{scriptsize}
The rows of the above matrix are linearly independent (given $\alpha\neq\epsilon$), and hence the minmax profiles have individual full rank for both players. 

We also need to verify that all pure strategy action profiles, which correspond to the extreme points of the payoff set $\mathcal{F}^\dagger$, have pairwise full rank, i.e., satisfy condition (C5). We do so for $\mathbf{r}=(0,0,0,\ldots, 1)$ for firms 1 and N; the remaining profiles can be verified similarly. For $\mathbf{r}=(0,0,0,\ldots, 1)$, the matrix $A_{1N}(\mathbf{r}):=[A_{1}(r_N=1); A_{N}(r_1=0)]$ is given by: 
\begin{scriptsize}
\[
\begin{blockarray}{cccc}
\begin{block}{[cccc]}
                   \alpha^{N-1}\cdot\epsilon
                  &   (1-\alpha)\cdot\alpha^{N-2}\cdot\epsilon 
                                  &  \ldots  
                  & (1-\alpha)^{N-1}\cdot(1-\epsilon)\\
                   \epsilon\cdot\alpha^{N-2}\cdot\epsilon
                  & (1-\epsilon)\cdot\alpha^{N-2}\cdot\epsilon  
		                    & \ldots 
                  & (1-\epsilon)\cdot(1-\alpha)^{N-2}\cdot(1-\epsilon)\\
                   \alpha^{N-1}\cdot\alpha
                  & (1-\alpha)\cdot\alpha^{N-2}\cdot\alpha 
		                   & \ldots  
                  &(1-\alpha)^{N-1}\cdot(1-\alpha)\\
                   \alpha^{N-1}\cdot\epsilon
                  & (1-\alpha)\cdot\alpha^{N-2}\cdot\epsilon  
 & \ldots 
                  & (1-\alpha)^{N-1}\cdot(1-\epsilon)\\                  
\end{block}
\end{blockarray}
\]
\end{scriptsize}

Here, the rows correspond to $r_1=0, r_1=1, r_N=0, r_N=1$, respectively, and the columns correspond to profiles of beliefs $\mathbf{b}$ ordered by the decimal value of their strings. Note that as the original profile has $r_1=0$ and $r_N=1$, the first and last rows above are indeed the same. It is then straightforward to verify that the first three rows are linearly independent, and therefore the matrix has row rank 3 as required by (C5). 
A similar procedure shows that the remaining pure action profiles also satisfy (C5) for any pair of firms.

}

\section*{Acknowledgment}
This material is based on research sponsored by the Department of Homeland Security (DHS) Science and Technology Directorate, Homeland Security Advanced Research Projects Agency (HSARPA), Cyber Security Division (DHS S\&T/HSARPA/CSD), BAA 11-02 via contract number HSHQDC-13-C-B0015.

\bibliographystyle{IEEEtran}
\bibliography{info-sharing}

\end{document}